%% file: hqetF.tex
\documentclass[11pt]{article}
\usepackage{epsfig}
\usepackage{amsfonts}
\usepackage{amsmath}

\newcommand{\la}[1]{\label{#1}}
\newcommand{\be}{\begin{equation}}
\newcommand{\ee}{\end{equation}}
\newcommand{\ba}{\begin{eqnarray}}
\newcommand{\ea}{\end{eqnarray}}
\newcommand{\bi}{\begin{itemize}}
\newcommand{\ei}{\end{itemize}}

\newcommand{\tr}{{\rm Tr\,}}

\renewcommand{\vec}[1]{{\bf #1}}

\newcommand{\RR}{{\rm I\kern -.2em  R}}

\def\lsi{\raise0.3ex\hbox{$<$\kern-0.75em\raise-1.1ex\hbox{$\sim$}}}
\def\gsi{\raise0.3ex\hbox{$>$\kern-0.75em\raise-1.1ex\hbox{$\sim$}}}

\newcommand{\hide}[1]{ }

\makeatletter \@addtoreset{equation}{section} \makeatother
\renewcommand{\theequation}{\arabic{section}.\arabic{equation}}
\makeatletter
\renewcommand\section{\@startsection {section}{1}{\z@}%
                                   {-5.5ex \@plus -1ex \@minus -.2ex}
                                   {2.3ex \@plus.2ex}%
                                   {\normalfont\large\bfseries}}
\renewcommand\subsection{\@startsection{subsection}{2}{\z@}%
                                     {-3.25ex\@plus -1ex \@minus -.2ex}%
                                     {1.5ex \@plus .2ex}%
                                     {\normalfont\normalsize\bfseries}}
\renewcommand\thesection {\@arabic\c@section}
\renewcommand\thesubsection   {\thesection.\@arabic\c@subsection}
\renewcommand{\@seccntformat}[1]{%
\csname the#1\endcsname.\hspace{1.0em}}
\makeatother

\begin{document}

\begin{titlepage}
\vspace{-1cm}
\begin{flushright}
\begin{tabular}{ll}
FTUV-09-1014 \\
IFIC/09-43\\
CERN-PH-TH-2009-170 \\
\end{tabular}
\end{flushright}

\vspace{2cm}
\begin{centering}
\vfill

{\Large\bf Heavy-light mesons in the $\epsilon$-regime}

\vspace*{0.8cm}

F.~Bernardoni$^{\rm a}$, 
P.~Hern\'andez$^{\rm a}$, 
S.~Necco$^{\rm b}$ 
\vspace*{0.5cm}

{\em $^{\rm a}$%
Dpto.\ F\'{\i}sica Te\'orica, 
Universidad de Valencia and IFIC-CSIC \\ 
Apt.\ 22085, E-46071 Valencia, Spain\\}

\vspace{0.3cm}

{\em $^{\rm b}$%
Theory Division, CERN \\
1211 Geneve 23, Switzerland}

\vspace*{2cm}

{\bf Abstract}
 
\end{centering}
 
\vspace*{0.4cm}

\noindent

We study the finite-size scaling of heavy-light mesons in the static limit. We compute two-point functions of chiral current densities as well as pseudoscalar densities in the $\epsilon$-regime of heavy meson Chiral Perturbation Theory  (HMChPT). As expected, finite volume dependence turns out to be significant in this regime and can be predicted in the effective theory in terms of the infinite-volume low-energy couplings. These results might be relevant for extraction of heavy-meson properties from lattice simulations.

\vspace{6cm} 
\end{titlepage}


\input{introduction.tex}

\input{mixed_int.tex}

\input{mixed_res.tex}

\input{hqet_int.tex}
\input{hqet_res.tex}

\input{matching.tex}

\input{concl.tex}
\input{appendices.tex}

\section*{Acknowledgements}

F.B. acknowledges the financial support of the FPU grant AP2005-5201. This work was  partially supported by the Spanish CICYT projects (FPA2006-60323, HA2008-0057 and CSD2007-00042), by the Generalitat Valenciana (PROMETEO/2009/116),  and by the European project FLAVIAnet (MRTN-CT-2006-035482). We thank M. Della Morte for useful comments.


\input hqetF.bbl
\end{document}

%% file: introduction.tex
\section{Introduction}

The simulations of heavy-light mesons made out of a heavy quark (charm or bottom) and a light one (up, down or strange) on the lattice are challenging because they require very large volumes in order to keep systematic errors under control. The reason is that the dynamics of these systems involve very distinct energy scales: the heavy-light ($hl$) meson mass, $M_{hl}$, the light pion masses $M_{ll}$ and $\Lambda_{QCD}$, that should all be kept sufficiently below the UV cutoff (i.e. the inverse lattice spacing), and sufficiently above the infrared one (i.e. the lattice box size). Both requirements can only be met in very large lattices. 

If the heavy quark mass is sufficiently large a good effective description is provided by heavy quark effective theory (HQET) 
\cite{Grinstein:1990mj,Eichten:1989zv,Georgi:1990um}, which is obtained in the limit of infinite heavy quark mass, or static limit. In this limit, the scale $M_{hl}$ disappears from the problem and the UV cutoff can in principle be as low as the cutoff used to describe light meson dynamics. Indeed this approximation has been extensively used to simulate heavy-light  mesons in lattice QCD (for a recent review on heavy flavour phenomenology from lattice QCD see \cite{Gamiz:2008iv}). 

Whether the heavy quark is treated in the static limit or not, an obvious question is if we can do better concerning the constraint on the box-size. After all, the finite-size scaling of heavy-light systems should be dominated by light pions physics, since these are the lightest modes in QCD. To the extent that pion physics can be described by chiral perturbation theory (ChPT) , it is conceivable that finite-size scaling of heavy-light systems can be accurately predicted using ChPT, as the finite-size scaling of light mesons is \cite{Gasser:1986vb,Gasser:1987ah,Gasser:1987zq}. 

In this paper, we investigate the possibility to predict the finite-size scaling of heavy-light systems, 
when the lightest pions are light compared to the inverse box size, from chiral perturbation theory. We will consider this problem in two limiting situations depending on the mass of the heavy quark:
\begin{itemize}
\item The heavy quark is significantly above the light one, but still treatable in ChPT: this would correspond to considering hl mesons in the the mixed-regime introduced in \cite{mixed}.
\item The heavy quark is static and therefore chiral dynamics can be treated in Heavy Meson Chiral Perturbation Theory (HMChPT): this would correspond to considering hl static mesons in the $\epsilon$-regime.  
\end{itemize} 
Even though these two situations are physically very different, the pion dynamics responsible 
for the finite-size scaling properties should be pretty much the same. It is therefore interesting to see explicitly how a quantitative matching of the finite-size effects takes place, by comparing the finite volume dependence of correlation functions in ChPT and HMChPT. 

We consider  the two-point function of  left-handed current densities that will be computed to next-to-leading order in the $\epsilon$-regime  in both effective theories. We will also consider the two-point correlator of pseudoscalar densities to the leading order, since finite-size effects are important already at this order. Anticipating the possible use of these results in simulations we also present the results in the partially-quenched (PQ) case.

The paper is organised as follows. In section~\ref{sec:mixed} we present the  results for the two-point functions in the mixed-regime of ChPT, when the heavy quark is treated in the $p$-regime and the light ones in the $\epsilon$-regime, that is in the so called mixed-regime.  In section~Ê\ref{sec:hqet} we discuss the formulation of HMChPT in the $\epsilon$-regime and present the results for the same correlators. 
In section~\ref{sec:comp} we compare both results and discuss the implications. In section~\ref{sec:lat} we briefly comment on the applications to lattice QCD and conclude in section ~\ref{sec:conclu}.

%% file: mixed_int.tex
\section{Heavy-light mesons in the mixed-regime of ChPT}
\label{sec:mixed}
The goal of this section is to study the finite-size scaling of heavy-light mesons in ChPT, when the light quarks are in the $\epsilon$-regime. We assume that the meson is composed of a heavy quark of mass $m_h$ and a light quark of mass $m_l$ and that both masses are very different $m_l \ll m_h$, but both can still be treated in the context of ChPT, that is 
\begin{eqnarray}
M^2_{xy} \equiv \frac{(m_x+m_y) \Sigma}{F^2} \ll (4 \pi F)^2, \;\;\; x,y = h, l.
\end{eqnarray}
Under this hypothesis the finite-size effects at NLO are predictable by using the common ChPT Lagrangian, that is:
\be \label{L_chir}
{\cal L}_{ChPT}=\frac{F^2}{4} \tr \left[ \partial_{\mu }U \partial_{\mu}U^{\dagger} \right]-\frac{\Sigma}{2}\tr \left[ {\cal M}^\dagger U  +U^{\dagger} {\cal M} \right]\,\,,
\ee
plus the counterterms one has to consider at one loop that were found by Gasser and Leutwyler and are proportional to the Low Energy Couplings (LECs) $L_i$.\cite{GL2}. 
The pseudo Nambu-Goldstone bosons are parametrised by $U\in SU(N)$, with $N=N_l+N_h$, being $N_l$ ($N_h$) the number of light (heavy) quarks.
We have absorbed the vacuum angle $\theta$ in the light quark masses. That is, the mass matrix ${\cal M}$ is:
\begin{equation}
{\cal M} \equiv \{m_1 e^{\frac{i \theta}{N_l}},\dots,m_{N_l}e^{\frac{i \theta}{N_l}},m_{N_l+1},\dots, m_N   \}  \,\,.
\end{equation}
The mesons are placed in a box of volume $V=L^3 T$, which  is sufficiently large to contain the typical QCD scale, and the heavy meson mass scale, but small compared to the lightest pion mass:
\begin{equation}
M_{hl} L \gg 1, \;\;M_{ll} L  \leq 1 . 
\end{equation}
In this situation it is expected that the finite volume effects associated to the scale $M_{hl} $ are exponentially suppressed, while those associated to $ M_{ll} $ are not. This regime of ChPT has been named {\it mixed-regime} in \cite{mixed,mixed2}, 
since some of the quarks are in the $p$-regime and some in the $ \epsilon $-regime. A convenient power-counting for the quark mass and momentum in this situation is
\begin{equation}
m_l \sim \epsilon^4, \;\; m_h \sim \epsilon^2, \;\; L^{-1} \sim p \sim \epsilon \,,
\label{eq:pc} 
\end{equation}
so that the LO mass of the heavy-light mesons will be:
\begin{eqnarray}
M^2_{h} \equiv \frac{m_h \Sigma}{F^2} \sim \epsilon^2\,\, .
\label{eq:mh}
\end{eqnarray}

We refer to \cite{mixed,mixed2} for further details on the implementation of ChPT in the mixed-regime, both in the full and partially-quenched theories. We just remind here that in this regime it is convenient to parametrise the pion field $U$ like:
\begin{equation}\label{parU}
U=\left(
\begin{array}{cc}
U_0 & 0 \\
0      & 1
\end{array}
\right)e^{\frac{2 i \xi}{F}},
\end{equation}
with the perturbative pion field $\xi$ satisfying the condition:
\begin{equation}\label{eq_xi}
\int d^4x \tr [T^a \xi(x)]=0 \quad \mbox{if}\quad T^a \in SU(N_l) ,\,\,\;\;\;a=1,\cdots,N_l^2-1.
\end{equation}
In this way, the light zero modes  to be treated non perturbatively are collected in $U_0$, and consequently they are dropped from $\xi$. In the references \cite{mixed, mixed2} two different parametrisations where used. We have explicitly tested the results with both of them. 

In refs.~\cite{mixed, mixed2}, the light-light meson correlators were computed. Here we extend this computation to the two-point correlation functions of heavy-light left-handed currents and  pseudoscalar densities, to relative ${\mathcal O}(\epsilon^2)$ order:
\begin{eqnarray}
  \tr[T^aT^b]{C_J}(t) & \equiv & 
 \int \! {\rm d}^3 x\, 
 \Bigl\langle {J}^a_0(x) {J}^b_0(0) \Bigr\rangle  \label{JLJL}\\ 
    \tr[T^aT^b]{C_P}(t) & \equiv & 
 \int \! {\rm d}^3 x\, 
 \Bigl\langle P^a(x) {P}^b(0) \Bigr\rangle ,
\end{eqnarray}
where in QCD the current and pseudoscalar densities can be formally defined as
\be
 J^a_\mu \equiv \bar\psi T^a \gamma_\mu P_- \psi, \;\;\;  P^a \equiv i\bar\psi T^a \gamma_5 \psi , \;\;\; 
 \; \la{Jamu}
\ee
$t$ represents the Euclidean time, $x=(\vec{x},t)$, and 
$P_- \equiv (1-\gamma_5)/2$. 
In order to represent a heavy-light meson, $T^a$ is any traceless generator with one index in the light subsector and the other one in the heavy one, for example:
\begin{eqnarray}
\left(T^a\right)_{ij} = {1\over 2} ( \delta_{ih} \delta_{jl} + \delta_{il} \delta_{jh}).  
\end{eqnarray}

As usual, in ChPT these operators can be represented \footnote{In order to simplify notation, we use throughout the same notation for the operators in QCD and their representation in ChPT.} to leading order in the momentum expansion by
\ba
J^a_\mu = 
 \frac{F^2}{2} \tr \Bigl[ T^a U \partial_\mu U^\dagger \Bigr],  \;\;  \;\;\; P^a =  i\frac{\Sigma}{2} \tr \Bigl[ T^a \left(U -U^\dagger \right)\Bigr] .
\ea

These results are useful in their own right to describe for example kaon correlators in a finite volume, when the $s$ quark is in the $p$-regime and the $u$ and $d$ are in the $\epsilon$-regime. We will also be interested in isolating the finite volume effects that survive in the static limit $m_h \rightarrow \infty$, which should match those obtained in HMChPT.  In order to recover the results for 
various full and partially-quenched situations of interest we consider the following computations:
\begin{itemize}
\item \emph{Case A}. \\
Degenerate heavy quarks: the Goldstone manifold is $SU(N_h+N_l)$, with $N_h$ quarks of mass $m_h$ and $N_l$ quarks with masses $m_{l}$ ($l=1,\cdots,N_l)$, with the counting rules of Eq.~(\ref{eq:pc}). We can consider then the quenched limit of the heavy quarks by taking the replica limit $N_h \rightarrow 0$. These results should match in the $m_h \rightarrow \infty$ limit those of HMChPT, where the heavy quarks are treated as static sources and all the light quarks are in the $\epsilon$-regime.
\item \emph{Case B}.\\
 Non-degenerate heavy quarks: the Goldstone manifold is 
$SU(N_h + N_s +N_l)$, with $N_h \rightarrow 0$ quarks of mass $m_h$ (i.e. the valence heavy quark), $N_s$ sea quarks of mass $m_s$ and $N_l$ of masses $m_{l_i}$, where both $m_h \sim m_s \sim \epsilon^2$. 
This can be matched to HMChPT in the limit $m_h \rightarrow \infty$.
This situation corresponds to having sea quarks both in the $\epsilon$ and in the $p$-regimes, for example if one considers $B$ or $D$ mesons in 2+1 dynamical simulations, where the $s$ quark is in the $p$-regime and 
the $u$ and $d$ quarks are in the $\epsilon$-regime. 
We can also quench the light quarks $N_l \rightarrow 0$ (quenching the heavy sea quarks $N_s \rightarrow 0 $ is equivalent to Case A),
 which would then correspond to the study of $D$ and $B$ mesons in a PQ mixed-action approach with sea quarks in the $p$-regime and the valence light quark in the $\epsilon$-regime. 
\end{itemize}

\subsection{Conventions}

We describe in the following our conventions for the propagators that we use to write down the results in a compact form.\\
The propagator for a pion with mass $M$ is, in finite volume:
\be\label{def_G}
G(x, M) \equiv  \frac{1}{V}\sum_{p}\frac{e^{ipx}}{p^2+M^2}\,\,.
\ee
Since some zero modes are factorised in the mixed regime, we also need to consider propagators in which they have been subtracted:
\be\label{def_Gbar}
\overline{G}(x,M) \equiv  \frac{1}{V}\sum_{p\ne 0}\frac{e^{ipx}}{p^2+M^2} \,\,.
\ee
The singlet part of the propagator gives rise to the following functions:
\ba
E(x,N_s,N_l, M)  \equiv  \frac{1}{V}\sum_{p\neq 0}\frac{e^{ipx}}{(p^2)^2F(p,N_s,N_l,M)}-\frac{N_s}{N_l^2VM^2}, \nonumber\\
 \;\; \Box E(x, N_s,N_l,M)  =-\frac{1}{V}\sum_{p\neq 0}\frac{e^{ipx}}{(p^2) F(p, N_s,N_l, M)}\,\,,
\ea
with $\Box =\partial_{x_\mu}\partial_{x_\mu}$, and
\begin{equation}
F(p,N_s,N_l,M)\equiv\frac{N_s}{p^2+M^2}+\frac{N_l}{p^2}.
\end{equation}
Once we integrate over space, the correlators exhibit exponential decay at large distances. This is represented by the function:
\be\label{def_P}
P(t, M) \equiv \int d^3\vec{x}G(x, M) ={\cosh \left[ M \left( {T \over 2} - |t| \right) \right] \over 2 M \sinh\left[ { M T \over 2}\right]},
\ee
when the pion running in the line has a mass of order $\epsilon^2$, or by:
\be\label{h1}
Th_1\left(\frac{t}{T} \right) \equiv \int d^3\vec{x}\,\overline{G}(x, 0) =\frac{T}{2}\left[\left(\frac{|t|}{T}-\frac{1}{2} \right)^2-\frac{1}{12} \right],
\ee
if the mass is of order $\epsilon^4$.\\
When two mesons propagate we need to introduce the function:
\begin{multline}
{k_{00}(M_1,M_2,t)} \equiv
\frac{1}{2}\sum_{\vec{p}}\Bigg\{2\frac{dP}{dt}(t,M_{1\vec{p}})\frac{dP}{dt}(t,M_{2\vec{p}}) \nonumber\\
- \left(P(t,M_{1\vec{p}})\frac{d^2P}{dt^2}(t,M_{2\vec{p}}) + (M_1 \leftrightarrow M_2) \right) \Bigg\}, \\    
\end{multline}
where we have introduced the shorthand $M_{a\vec{p}}\equiv \sqrt{M_a^2+\vec{p}^2}$. This expression is substituted by:
\be
{\overline{k}_{00}(M_1,t)} \equiv  \lim_{M_2\rightarrow 0} \left(k_{00}(M_1,M_2,t)+\frac{P(t,M_1)M_1^2}{2TM_2^2}\right),
\ee
when $M_2$ lays in the $\epsilon$-regime.

%% file: mixed_res.tex
\subsection{Left-current correlator}

\vspace{0.5cm}
{\it Case A}

The result for the left-correlator at NLO using the above definitions is:
\begin{multline}
\label{eq:caj}
C^{(A)}_J(t)= \frac{F_{(A)}^2}{2}M_{(A)}^2P(t,M_{(A)})\\
-\frac{T}{2V}\Bigg\{\left(N_h-\frac{1}{N_h}\right) {k_{00}(M_{h},M_{hh},t)}+ \left(\frac{1}{N_h}+\frac{1}{N_l} \right)  {k_{00}(M_{h},M_{\eta_h},t)}    \\
+\left(N_l-\frac{1}{N_l}  \right){\overline{k}_{00}(M_{h},t)}\Bigg\},
\end{multline}
where we have defined 
\begin{eqnarray}
&&F_{(A)}^2 \equiv F^2-\frac{1}{2}\left( N_h-\frac{1}{N_h}  \right)G\left(0,M_{hh}\right)-\frac{N_l}{2}\overline{G}\left(0,0\right)-\frac{N_l+N_h}{2}G\left(0,M_{h}\right) \label{FANLO}\nonumber \\
&&-\left(\frac{1}{2(N_l+N_h)}+\frac{1}{2N_h} \right)G\left(0,M_{\eta_h}\right)+\frac{1}{2}E(0,N_h,N_l,M_{hh})+8M_{h}^2(2 L_4N_h+L_5) ,\\
&&M_{(A)}^2 \equiv M_h^2\left[1-\frac{1}{F^2}\left(8M_h^2(2L_4N_h+L_5-4N_hL_6-2L_8) -\frac{2N_h+3N_l}{3(N_l+N_h)^2}G(0,M_{\eta_h})\right. \right. \nonumber \\
&&\left.\left.+\frac{1}{6}\frac{\Box E(0,N_h,N_l,M_{hh})}{M_h^2}  \right) -\frac{2}{\mu_h}\left(\frac{1}{6N_l}-\frac{N_l}{4}-\frac{\mu_{l}}{4}\langle (U_0+U_0^{\dagger})_{l l} \rangle  \right) \right] \,\,,\label{MANLO}
\end{eqnarray}
and 
\begin{eqnarray}
\mu_i &\equiv & m_i\Sigma V, \\
M_{\eta_h}^2 & \equiv & \frac{N_l}{N_l+N_h}M_{hh}^2,
\end{eqnarray}
while $M^2_h$ is defined in Eq.~(\ref{eq:mh}).

A few observations are in order. The UV divergences in  $F^2_{(A)}$ and in $M_{(A)}^2$ can be shown to cancel in the renormalisation of the NLO couplings of Gasser and Leutwyler, $L_i's$. We have also checked that the result matches the result of \cite{mixed} for non-degenerate quarks in the $\epsilon$-regime in the appropriate limit.  

This result represents the finite-size scaling of kaon-like states ($m_h=m_s$ and $m_l =m_u=m_d$) in the mixed-regime for various situations:
\begin{itemize}
\item $2+1$ dynamical simulations setting:  $N_h=1, N_l= 2$, 
\item PQ simulations where the $h$ quarks are quenched and the $l$ quarks are dynamical by taking the replica limit $N_hÊ\rightarrow 0$ of Eq.~(\ref{eq:caj}),
\item PQ simulations where the $l$ quarks are all quenched  or partially quenched, while the $h$ quarks are dynamical. In this case, the appropriate value of $N_l$ must be taken, but also the zero-mode integrals  $\langle (U_0+U_0^{\dagger})_{ll}\rangle$ need to be properly defined \footnote{Note that one cannot consider a fully quenched theory with $N_h=N_l=0$ on the basis of Eq.~(\ref{eq:caj}), because the singlet has been integrated out\cite{Damgaard:2001js}.}.
\end{itemize}
We discuss now the result of the zero-modes integrals $\langle (U_0+U_0^{\dagger})_{ll}\rangle $ (for further details see \cite{Damgaard:2007ep,mixed2}).  In order to treat the situation where some light quarks might be quenched, we distinguish within the light  ($\epsilon$-regime) sector $N_l$ sea quarks and $N_q$ quenched ones.
When restricting to a topological sector $\nu$, the averages in all the cases described above can be obtained in a compact and general form, in terms of the partition functional \cite{Kanzieper:2002ix,Splittorff:2002eb}:
\begin{equation}
\label{eq:zero-mode}
\mathcal{Z}^\nu_{N_q,N_l+N_q}(\{\mu_i\})
=
\frac{\det[\mu_i^{j-1}\mathcal{J}_{\nu +j-1}(\mu_i)]_{i,j=1,\cdots 2N_q+N_l}}
{\prod_{j>i \ge 1}^{N_q}(\mu_j^2-\mu_i^2)\prod_{j>i\ge N_q+1}^{2N_q+N_l}(\mu_j^2-\mu_i^2)},
\end{equation}
and its derivatives.
Here $\mathcal{J}$'s are defined as
$\mathcal{J}_{\nu+j-1}(\mu_i)\equiv (-1)^{j-1} K_{\nu+j-1}(\mu_i)$ 
for $i=1,\cdots N_q$ and 
$\mathcal{J}_{\nu+j-1}(\mu_i)\equiv I_{\nu+j-1}(\mu_i)$ 
for $i=N_q+1,\cdots 2N_q+N_l$, 
where $I_\nu$ and $K_\nu$ are the modified Bessel functions. 
For the observable of interest here, the result in the  theory with all the light quarks dynamical is:
\be
\frac{\langle (U_0+U_0^{\dagger})_{ll} \rangle_{\nu}}{2}
~\equiv~
\frac{\partial}{\partial \mu_{l}} 
\ln \mathcal{Z}^\nu_{0,N_l+0}(\{\mu_l\})\,\,,
\ee
while in theories where the valence light quark is quenched  is:
\begin{eqnarray}
\label{eq:conden}
\frac{\langle (U_0^{\dagger}+U_0)_{v v}\rangle_{\nu}}{2}&\equiv&
\lim_{\mu'_{v} \to \mu_{v}}\frac{\partial}{\partial \mu'_{v}} 
\ln \mathcal{Z}^\nu_{1,1+N_l}(\mu_v,\mu'_v, \{\mu_{l}\}).
\end{eqnarray}

\vspace{0.5cm}
{\it Case B}

In this case we will denote respectively the squared mass of the heavy-light mesons and the decay constant at NLO by $M_{(B)}^2$ and $F_{(B)}$. The result in case B in the replica limit $N_h \rightarrow 0$ is:
\begin{multline}
\label{eq:cbj}
C^{(B)}_J(t)= \frac{F_{(B)}^2}{2}M_{(B)}^2P(t,M_{(B)})\\
-\frac{T}{2V}\Bigg\{\left(N_l-\frac{1}{N_l}\right){\overline{k}_{00}(M_h,t)}
+\frac{M_{hh}^4N_sN}{N_l(M_{hh}^2N-M_{ss}^2N_l)^2}{{k}_{00}(M_h,M_{\eta_s},t)}\\
+\frac{M_{ss}^4N_l+M_{hh}^4N-2M_{hh}^2M_{ss}^2N}{(M_{hh}^2N-M_{ss}^2N_l)^2}{{k}_{00}(M_h,M_{hh},t)}
+N_s{{k}_{00}(M_{s}, M_{hs},t)}\\
-\frac{M_{hh}^2(M_{hh}^2-M_{ss}^2)}{(M_{hh}^2N-M_{ss}^2N_l)}\left({\frac{d}{dM_2^2}{k}_{00}(M_{h},M_2,t)}\right)_{M_2=M_{hh}}\Bigg\}\,\,,
\end{multline}
where we have defined the shorthand 
\begin{equation}\label{etas}
M_{\eta_s}^2=\frac{N_l}{N_l+N_s}M_{ss}^2, 
\end{equation}
 and $N= N_s+N_l$, while
\ba
F_{(B)}^2 &=&F^2 -\frac{N_s}{2}G\left(0,M_{hs}\right)-\frac{N_s}{2}G\left(0,M_s \right)-\frac{N_l}{2} \overline{G}(0,0)-\frac{N_l}{2}G(0,M_h)+\frac{1}{2}E^{\epsilon}(0,N_s,N_l,M_{ss})\label{FBNLO} \nonumber \\
&& +8(M_h^2 L_5+N_sL_4M_{ss}^2)+\frac{1}{2}\left(\frac{N_s M_{ss}^2M_{hh}^2}{(NM_{hh}^2-N_lM_{ss}^2)^2}-\frac{M_{hh}^2-M_{ss}^2}{NM_{hh}^2-N_lM_{ss}^2} \right)G(0,M_{hh}) \nonumber \\
&&-\left( \frac{1}{2N}+\frac{N_sM_{ss}^2M_h^2}{(NM_{hh}^2-N_lM_{ss}^2)^2}-\frac{M_{hh}^2-M_{ss}^2}{2(NM_{hh}^2-N_lM_{ss}^2)} \right)G(0,M_{\eta_s})\nonumber \\
&& -\frac{M_h^2(M_{ss}^2-M_{hh}^2)}{NM_{hh}^2-N_lM_{ss}^2}\frac{d}{dM_{hh}^2}G(0,M_{hh}),\\
M_{(B)}^2 &=& M_h^2\left[1-\frac{1}{F^2}\left(8M_{ss}^2N_s(L_4-2L_6)
+4M_{hh}^2(L_5-2L_8) -\frac{(M_{hh}^2-M_{ss}^2)}{NM_{hh}^2-N_lM_{ss}^2}G(0,M_{hh}) \right. \right. \nonumber \\
&& \left. \left.+\frac{1}{6}\frac{\Box E^\epsilon(0,N_s,N_l,M_{ss})}{M_h^2}+\left(\frac{M_{ss}^2N_s}{6N^2M_h^2}-\frac{1}{N}+\frac{M_{hh}^2-M_{ss}^2}{NM_{hh}^2-N_lM_{ss}^2}\right)G(0,M_{\eta_s}) \right)\right. \nonumber \\
&&\left. -\frac{2}{\mu_h}\left( \frac{1}{6N_l} -\frac{N_l}{4}-\frac{\mu_{{l}}}{4}\langle(U_0+U_0^\dagger)_{l l}\rangle  \right) \right].\label{MBNLO}
\ea

We have performed several consistency checks of these results. For $m_s = m_h$ Case A is recovered. UV divergences do cancel. We recall that $\overline{G}(0,0)$ has no divergences in dimensional regularisation, and it is given by \cite{Hasenfratz:1989pk}:
\begin{equation}\label{beta1}
\overline{G}(0,0)\equiv -\frac{\beta_1}{\sqrt{V}}\,,
\end{equation}
where $\beta_1$ is a so-called shape coefficient, which depends on $T/L$.\\
In the replica limit $N_l \rightarrow 0$, this result represents the finite-size scaling of kaon-like correlators in PQ simulations where the $N_s$ sea quarks are  in the $p$-regime, while the light valence quarks are in the $\epsilon$-regime, a setup that might be useful  in mixed-action simulations. 

In either case A or B, we expect these predictions to match the ones of HMChPT in the  limit  $m_h \rightarrow \infty$, since this should recover a static limit of the valence quark. Indeed,   the leading volume dependence in $F_{(A)}$ and $F_{(B)}$ or $M_{(A)}$ and $M_{(B)}$ can be shown to be associated to the light sector only and therefore should be independent of the heavy mass scale. We will explicitly show how this happens in section \ref{sec:comp}. 

\subsection{Pseudoscalar correlator} 

Another interesting observable is the pseudoscalar density correlator as regards the finite volume dependence, because finite-size effects appear already at the leading-order as opposed to 
the correlator of the left current, where they appear first at NLO.

The result at LO in the chiral expansion is the same for cases A and B:
\begin{equation}
C_{P}(t)=  \frac{\Sigma^2}{2F^2}P(t,M_{h})\left[\langle  (U_0+U_0^\dagger)_{ll} \rangle +2\right] .
\label{pseudocorChPT}
\end{equation}
In this case, it is trivial to see that all the significant volume dependence comes from the zero-mode averages, which involve only the light sector.

%% file: hqet_int.tex
\section{Static heavy-light mesons in finite volume HMChPT}
\label{sec:hqet}

The effects of pion dynamics in the properties of static heavy-light mesons can be predicted in 
HMChPT \cite{Burdman:1992gh,Wise:1992hn,Yan:1992gz}. 
Most calculations of chiral corrections have been done in infinite volume. 
The authors of \cite{Arndt:2004bg} considered also chiral corrections in $B$ parameters of neutral $B$ meson mixing and heavy-light decay constants in a finite volume, but in the $p$-regime. 
We want to go further into the chiral limit by considering the $\epsilon$-regime for the light quarks. As far as we know, this regime has not yet been explored in HMChPT. However part of the technology we have used was developed in \cite{Smigielski:2007pe} to perform $\epsilon$-regime calculations in baryon ChPT.
We present our NLO results for left-current correlators in the $p$-, $\epsilon$- and mixed regimes, and LO results for pseudoscalar density correlator.

\subsection{Formulation and conventions}\label{subsec:hqet_int}

In the limit in which the mass of the heavy quark $m_h$ goes to infinity it is expected that QCD simplifies. For example the interactions among the quark and the antiquark in a meson become spin independent, and if we consider processes in which only low momenta are involved, the heavy antiquark (or quark, if one prefers) can be decoupled. An effective field theory to analyse this situation can be built by rewriting the heavy-quark momentum $p_{\mu} $ as: $p_{\mu}=m_hv_{\mu}+k_{\mu}$ and keeping only the leading term in the residual momentum $k_{\mu}/m_h$. 
To recover the peculiarities of QCD, for example the chromomagnetic interactions, one has to rewrite the QCD Lagrangian as a series in powers of $k_{\mu}/m_h$ and consider also those terms that vanish in the $m_h \to \infty $ limit, up to the required degree of precision.\\ 
In this work we have just considered the leading order in the above expansion. In such a case, the interactions with the pions are not able to modify the unitary velocity $v_{\mu}$ of the heavy-light mesons. 
We adopt a covariant representation, where the degenerate pseudoscalar and vector states are treated as a single field $H$ which
is usually labelled by $v$ and the flavour $a=1,\cdots,N_l$ of the light quark. In the Euclidean space we have 
\footnote{While the formulation in Minkowsky space can be exhaustively found in the standard literature (see e.g.\protect\cite{Manohar:2000dt,Casalbuoni:1996pg}), we find useful to start from the beginning with the formulation in the Euclidean space. Notice however that for $\vec{v}\neq \vec{0}$ the Euclidean formulation is problematic \protect\cite{Aglietti:1992in,Aglietti:1993hf}, and only the case $v=(\vec{0},i)$ will be considered.}
\begin{eqnarray}
H_v^a &=& \left( \frac{1-i v_{\rho}\gamma _{\rho}}{2}\right)[-iP_{\mu}^{a*}\gamma_{\mu}-iP^a \gamma_5], \label{h1v}\\
\overline{H}_v^a &=&[-iP_{\mu}^{a*\dagger}\gamma_{\mu}-iP^{a\dagger} \gamma_5]  \left( \frac{1-i v_{\rho}\gamma _{\rho}}{2}\right),\label{h2v}
\end{eqnarray}
where $P^*$ and $P$ represent respectively the vector and the pseudoscalar mesons, and $P^*$ satisfies:
\begin{equation}
v\cdot P^*=0 \,\,.
\end{equation}
The four-velocity $v=(\vec{v},v_4)$ satisfies the condition $v^2=-1$; the rest frame corresponds to $v=(\vec{0},i)$. 
We use the conventional HQET normalisation of the states
\begin{equation}
\langle H_v^a| H_{v'}^b    \rangle=2v_4 (2\pi)^3\delta_{vv'}\delta^{ab},
\end{equation}
according to which $H$ fields have mass dimension -3/2. For simplicity, we drop the $v$ label from here on. \\
The Euclidean Dirac matrices are chosen to be Hermitean,
\begin{equation}
\gamma_\mu^\dagger=\gamma_\mu,\;\;\;\gamma_5=\gamma_5^\dagger =\gamma_1\gamma_2\gamma_3\gamma_4,
\end{equation}
and satisfy the anticommuting relations
\begin{equation}
\left\{\gamma_\mu,\gamma_\nu  \right\}      =2\delta_{\mu\nu}.
\end{equation}
The projector $(1-i v_{\rho}\gamma _{\rho})/2$ in Eqs.(\ref{h1v},\ref{h2v}) retains only the particle component of the heavy quark.

In a theory with $N_l$ light quarks, and when dealing only with light mesons, one usually parametrises them with an $SU(N_l)$ matrix $U=\exp(2 i \xi/F)$. If
we rotate in flavour space the left (right) handed light quarks by a special unitary matrix L (R), the $U$ field will transform like $U \to
LUR^{\dagger}$. As it is well known, when dealing with heavy-light mesons it is convenient to use the field $\sqrt{U}$ to avoid that the parity transformation  involves the pseudo-Goldstone boson field \cite{Burdman:1992gh,Wise:1992hn,Yan:1992gz}.
$\sqrt{U}$ transforms like $\sqrt{U} \to L\sqrt{U} W^{\dagger}$ or $\sqrt{U} \to W\sqrt{U} R^{\dagger}$, where $W$ is a complicated function of $R$, $L$ and the meson field $\xi$. Then $H$ transforms as:
\begin{equation}
H\rightarrow HW^{\dagger} \,\,.
\end{equation}
To write more easily a chiral invariant Lagrangian we build combinations of $\xi$, that like $H$, only transform with $W$ or $W^{\dagger}$ under chiral rotations:
\begin{eqnarray}
{\cal V}_{\mu} & \equiv & \frac{i}{2} (\sqrt{U}^{\dagger}\partial_{\mu}\sqrt{U}+\sqrt{U} \partial_{\mu}\sqrt{U}^{\dagger}),\quad  \quad {\cal V}_{\mu}  \to  W {\cal V}_{\mu}W^{\dagger} +i W {\partial}_{\mu}W^{\dagger},\\
{\cal A}_{\mu} & \equiv & \frac{i}{2} (\sqrt{U}^{\dagger}\partial_{\mu}\sqrt{U}-\sqrt{U} \partial_{\mu}\sqrt{U}^{\dagger}),  \quad\quad {\cal A}_{\mu}  \to  W {\cal A}_{\mu}W^{\dagger} \,\,.
\end{eqnarray}
Then, at leading order in $1/m_h$, a Lagrangian that is both Lorentz and chiral invariant is:
\begin{equation}
\label{lagrangian}
{\cal L}^{(0)}_{HMChPT}=i \mbox{Tr}[\overline{H}^av_{\mu}(\partial_{\mu}\delta^{ab}+i{\cal V}^{ba}_{\mu})H^b]-ig_{\pi}
\mbox{Tr}[\overline{H}^aH^b\gamma_5\gamma_{\nu}{\cal A}^{ba}_{\nu}].
\end{equation}
The dynamics of the pseudo Nambu-Goldstone bosons is still given by the chiral Lagrangian in Eq. (\ref{L_chir}).\\
From the kinetic part of Eq.~(\ref{lagrangian}) one can extract the $P$ and $P^*$ propagators. For $v=(\vec{0},i)$ we obtain
\begin{eqnarray}
\langle P^a(x) P^{b\dagger}(y)\rangle &= & \delta^{ab}V(x-y)\\
\langle P_\mu^{a*}(x) P_\nu^{b*\dagger}(y)\rangle &= & \delta^{ab}V(x-y)(\delta_{\mu\nu}-\delta_{\mu 4}\delta_{\nu 4}),
\end{eqnarray}
where $V(x-y)=\frac{1}{2}\delta(\vec{x}-\vec{y})\theta(x_4-y_4)$. See App. \ref{appA1} for a more detailed discussion.\\
The term of the Lagrangian in Eq.~(\ref{lagrangian}) proportional to $g_\pi$ represents the interaction of $P$, $P^*$ 
with an odd number of pseudo-Goldstone bosons. In particular, by expanding 
\begin{equation}
\sqrt{U}=e^{i\xi/F}
\end{equation}
we obtain the $P^* P \xi$ and  $P^* P^* \xi$ couplings
\begin{equation}
{\cal L}^{(0)}_{HMChPT} = ...+  \frac{2ig_{\pi}}{F}\partial_{\nu }\xi^{ba}   \left(P^{a\dagger} P_{\nu}^{b*} -    P_{\nu}^{a*\dagger}P^b  \right)+
\frac{2g_{\pi}}{F}  \partial_{\nu }\xi^{ba}           P_{\alpha}^{a*\dagger}P_{\beta}^{b*} \epsilon_{\alpha \lambda\beta\nu}v_{\lambda}
\end{equation}
at leading order in the $1/m_h$ expansion. Note that the $PP\xi$ coupling vanishes because of parity. We adopt the convention
\begin{equation}
\epsilon_{1234}=1.
\end{equation}
There are several determinations of $g_\pi$ on the lattice, in the quenched case \cite{deDivitiis:1998kj,Abada:2002xe,Abada:2003un} and more recently in full QCD \cite{Negishi:2006sc,Ohki:2008py,Becirevic:2009yb}.

A number of operators can appear at next-to-leading order in the chiral expansion \cite{Boyd:1994pa}, however, if we omit contact terms, the only ones relevant to us are: 
\begin{equation}
\delta {\cal L}_{HMChPT}^{(2)}=-2\sigma_1 \mbox{Tr}[\overline{H} \widetilde{\cal M} H]-2\sigma_1' \mbox{Tr}[\overline{H} H] \mbox{Tr}[\widetilde{{\cal  M}}].
\end{equation}
where $\widetilde{\cal  M}$ has been defined as:
\begin{equation}
\widetilde{{\cal  M}}\equiv \frac{1}{2}(\sqrt{U} {\cal M}^{\dagger}\sqrt{U}+\sqrt{U}^{\dagger} {\cal M}\sqrt{U}^{\dagger}).
\end{equation}
The operator with the quantum numbers of the left current made of a heavy quark and a light antiquark with flavour index $l$, with the minimum power of $H$ fields derivatives and mass insertions is:
\begin{equation}
\label{leftcur}
{\cal J}^{l}_{\mu}\equiv \frac{a}{2}\mbox{Tr}[\gamma_{\mu} P_- (H \sqrt{U}^{\dagger})^l] \,\,.
\end{equation}
At leading order, the normalisation constant $a$ is related to the pseudoscalar meson decay constant $F_P$ and the corresponding mass $M_P$
 by the relation
\begin{equation}\label{def_a}
a=F_P \sqrt{2 M_P}.
\end{equation}
The vector meson decay constant is then given by
\begin{equation}
F_{P^*}=M_PF_P,
\end{equation}
while for the masses one has $M_{P^*}=M_P$.\\
To represent the left current at NLO additional terms appear \cite{Boyd:1994pa}:
\begin{equation}
\label{leftcurCT}
\delta {\cal J}^{l}_{\mu}=\frac{a\eta_0}{4}\mbox{Tr}[\gamma_{\mu}P_- (H  \widetilde{{\cal  M}}  \sqrt{U}^{\dagger})^l]+\frac{a\eta_3}{4}\mbox{Tr}[\gamma_{\mu} P_- (H   \sqrt{U}^{\dagger})^l]\mbox{Tr}[\widetilde{ {\cal  M}}]\,\,,
\end{equation}
that absorb the UV divergences.

In the static case, $v=(\vec{0},i)$ the heavy-light  left current correlator takes the form
\begin{equation}
  Q_{\mu \nu} {\cal C}_{J}^{l (I)}(t)  \equiv  
 \int \! {\rm d}^3 x\, 
 \Bigl\langle {\cal J}^l_{\mu}(x) \overline{{\cal J}}^{l}_{\nu}(0) \Bigr\rangle,  \qquad Q_{\mu \nu}\equiv (-\delta_{\mu \nu}+2\delta_{\mu 4}\delta_{\nu 4}) \;,
\end{equation}
where
\begin{equation}
\overline{{\cal J}}^{l}_{\mu}\equiv\frac{a}{2}\mbox{Tr}[\gamma_{\mu} P_- (\sqrt{U} \overline{H})^l] \,\,.
\end{equation}

Using this notation we isolate the time dependence in ${\cal C}_{J}^{l (I)}(t)$ for later comparison with the mixed-regime result. We will use the index $I=p$ to indicate the case where all light quarks are in the $p$-regime and $I=\epsilon$, where all are in the $\epsilon$-regime. 
Moreover, we will consider the case when some light quarks are in the $p$-, others are in the $\epsilon$-regime, and denote it by $I=m$. We are interested 
in the cases $I=\epsilon$ and $I=m$, with $\epsilon$-regime valence quarks,  which should match respectively the $m_h \rightarrow \infty$ limit of cases $A$ and $B$ in the ChPT computation.

Similarly,  at leading order in the momentum/mass expansion, the operator representing the pseudoscalar density is
\begin{eqnarray}
{\cal P}^l \equiv \frac{i a}{4} {\rm Tr} \left[\gamma_5 H^b \left(\sqrt{U}^{bl} + \sqrt{U}^{\dagger\;bl} \right)\right],
\end{eqnarray}
where $a$ is the normalisation factor defined in Eq. (\ref{def_a}).
In the case  where all light quarks are in the $\epsilon$-regime, we will give the LO result for the correlator
\begin{equation}
  {\cal C}_P^{l}(t)  \equiv  
 \int \! {\rm d}^3 x\, 
 \Bigl\langle {\cal P}^l(x) \overline{{\cal P}}^{l}(0) \Bigr\rangle,  
\end{equation}
with 
\begin{eqnarray}
\overline{{\cal P}}^{l} \equiv\frac{i a}{4} {\rm Tr} \left[\gamma_5 \overline{H}^b \left(\sqrt{U}^{bl} + \sqrt{U}^{\dagger\;bl} \right)\right].
\end{eqnarray}
Note that we are using calligraphic characters to denote quantities calculated in HMChPT to distinguish them from the corresponding quantities of ChPT.

%% file: hqet_res.tex
\section{HMChPT in $p$-regime}
We consider HMChPT with $N_l$ degenerate light quarks of mass $m$ lying in the $p$-regime. 
Making use of the space integrals given in App. \ref{app_a2} we obtain, for $t \ne 0$:
\begin{multline}
{\cal C}^{(p)l}_{J}(t)= \theta(t)\frac{a^2}{8} \exp\left(-\Delta M^{(p)} t\right)
\Bigg\{
1+2m(\eta_0+N_l\eta_3) \\
+\frac{1}{2F^2 L^2} \left(N_l-\frac{1}{N_l}\right)\frac{1}{L} \sum_{\vec{p}} \left[\left(P(t,M_{\vec{p}})-P(0,M_{\vec{p}})  \right) \left(1+g_{\pi}^2 \frac{\vec{p}^2}{M_{\vec{p}}^2}\right)   \right]  \Bigg\} \,,\\
\vspace{-1cm}
\end{multline}
with $M^2=2m\Sigma/F^2$ and $M_{\vec{p}}=\sqrt{M^2+\vec{p}^2}$, while 
\ba
\label{dmassp}
\Delta M^{(p)} \equiv 2 m (\sigma_1+ N_l \sigma'_1) +  g_\pi^2 \frac{M^2}{4 F^2 L^3} \left(N_l-\frac{1}{N_l}\right) \sum_{\vec{p}} \frac{1}{M_{\vec{p}}^2}.
\ea
 The function $P$ has been already defined in Eqs. (\ref{def_P}).\\

In dimensional regularisation  $\sum_{\vec{p}}P(0,M_{\vec{p}})$  and $\sum_{\vec{p}}P(0,M_{\vec{p}})M_{\vec{p}}^{-2}$ contain divergences, while $\sum_{\vec{p}}M_{\vec{p}}^{-2}$ is finite. To show this we rewrite:
\be
P(0,M_{\vec{p}})=\frac{1}{2M_{\vec{p}}}\left(1+\frac{2}{e^{M_{\vec{p}}T}-1} \right) \,\,,
\ee
and define in $s$ dimensions:
\be
G_{s,r}(0,M) \equiv \frac{1}{\prod_{i=1}^s L_i}\sum_p \frac{1}{(p^2+M^2)^r}\,\,.
\ee
where in our case $L_{1,2,3}=L, L_4=T$.
In the $\overline{MS}$ scheme we get:
\ba
 G_{4,1}(0,M)  &=& 2M^2 \lambda(\mu)+\frac{M^2}{(4\pi)^2}\ln \frac{M^2}{\mu^2}+ G^V_{4,1}(0,M), \\
G_{3,\frac{3}{2}}(0,M) &=& -8 \lambda(\mu)-\frac{1}{4\pi^2}\left(\ln \frac{M^2}{\mu^2}+1\right)+ G^V_{3,\frac{3}{2}}(0,M), \\
G_{3,1}(0,M) &=&-\frac{M}{4\pi}+ G^V_{3,1}(0,M). \\
\ea
In this expression $\lambda(\mu)$ contains the divergence,
\be
\lambda(\mu) \equiv  \frac{1}{16\pi^2}\mu^{4-d}\left[\frac{1}{d-4}-\frac{1}{2}(\ln(4\pi)+\Gamma'(1)+1)   \right],  
\ee
while $G^V_{n,r}$ contains the finite volume dependence, which can be expressed as a series of Bessel functions:
\ba
G^V_{4,1}(0,M)                        &\equiv & \frac{1}{4\pi^2}\sum_{n\ne 0}\frac{M}{|z|}K_{-1}(M |z|)\\
G^V_{3,\frac{3}{2}}(0,M) &\equiv & \frac{1}{2\pi^2} \sum_{n\ne 0} K_0(M |z|) \\
G^V_{3,1}(0,M)               &\equiv & \frac{1}{(2\pi)^{\frac{3}{2}}} \sum_{n\ne 0} \sqrt{\frac{M}{|z|}}K_{-\frac{1}{2}}(M |z|) \,
\ea
where $z=(n_1L_1,\cdots ,n_s L_s)$, $\{n_i \in \mathbb{Z}\; ;i=1,\cdots,s\}$. \\
It can be shown that:
\ba
  \frac{1}{L^3}\sum_{\vec{p}} P(0,M_{\vec{p}})&=& G_{4,1}(0,M), \nonumber\\
  \frac{1}{L^3}\sum_{\vec{p}} \frac{P(0,M_{\vec{p}})}{M_{\vec{p}}^{2}}&=& 
  \frac{1}{2} G_{3,\frac{3}{2}}(0,M) + \frac{1}{L^3}\sum_{\vec{p}}
  \frac{M_{\vec{p}}^{-3}}{e^{M_{\vec{p}}T}-1}.
\ea
Defining the renormalised coupling:
\begin{equation}
\eta_i=\eta_i^{(r)}+\overline{\eta_i}\lambda(\mu) \,\,,
\end{equation}
and requiring the cancellation of UV divergences we obtain, in agreement with \cite{Goity:1992tp}\footnote{Note that there is no standard convention for the normalisation of the couplings $\bar\eta_0$ and $\bar\eta_3$. }
\begin{equation}
\bar\eta_0+N_l\bar\eta_3=\frac{\Sigma}{F^4}\left(N_l-\frac{1}{N_l}  \right)(1+3g_\pi^2) \,.
\end{equation}
Obviously one can also reproduce the infinite volume result by taking the limits $T,\,L\to \infty$.

\section{HMChPT in $\epsilon$-regime}
\label{epsreg}

\subsection{Setup}
We consider now $N_l$ light quarks lying in the $\epsilon$-regime.
In this regime it is convenient to use the following parametrisation for the pseudo Nambu-Goldstone fields:
\begin{equation}
\label{param_e}
U=U_0e^{\frac{2i\xi}{F}}
\end{equation} 
for which the integration measure is known up to NLO \cite{Hansen:1990un} and gives no contribution to our observables. Here $\xi$ contains the non zero modes of the pions and is a perturbative field $\xi \sim \epsilon$. 

The complication, in heavy-light mesons calculations, is that we need to express $\sqrt{U}$ as a function of $\sqrt{U_0}$ and $\xi$, up to $\epsilon ^2$ corrections. The solution can be written in the form:
\begin{equation}
\sqrt{U}=\sqrt{U_0} \left(1+\frac{i A}{F}-\frac{B}{2F^2} \right)+O(\epsilon^3)
\end{equation}
where $A$ and $B$ are Hermitian matrices ($A$ is also traceless), respectively of order $\epsilon$ and $\epsilon ^2$, linear and quadratic in the components of $\xi$.
Imposing:
\begin{equation}
\left(\sqrt{U}\right)^2 = U+O(\epsilon^3)
\end{equation}
we obtain the system of $2N_l^2-1$ equations:
\begin{eqnarray}
A+\sqrt{U_0}^{\dagger}A \sqrt{U_0} &=& 2 \sqrt{U} \\
B+\sqrt{U_0}^{\dagger}B \sqrt{U_0} &=& 4 \sqrt{U}^2-4\sqrt{U} A+2A^2
\end{eqnarray}
which can be solved in a particular system of coordinates for $U_0$. 

We have not found a simple way to solve the equations for general $N_l$, so we have considered the particular case of $SU(2)$, that is $N_l =2$. One convenient choice for this group is to use the hyperspherical coordinates:
\begin{equation}
\sqrt{U_0}= \cos \psi +i \sin \psi \sin \theta \cos \phi {\sigma_1}+i\sin \psi \sin \theta \sin \phi {\sigma_2}+i \sin \psi
\cos \theta  {\sigma_3} \,\,,
\end{equation}
where $\sigma_i$ are the Pauli matrices and the angle ranges are:
\begin{equation}
\label{range}
\psi \in [0, \pi], \quad \theta \in [0, \pi], \quad \phi \in [0, 2 \pi] \,\,.
\end{equation}
Note that to parametrise $U_0$ we just need to extend the range of $\psi$: $\psi \in [0, 2\pi]$.

As usual, it is worth to perform the contractions of the non zero modes first and then perform the non perturbative integrations of $\psi$, $\phi$  and $\theta$ over the range specified by (\ref{range}).\\
The Haar integration measure to be used for the zero modes is, in hyperspherical coordinates:
\begin{equation} 
\int [{\cal D}U_0]=\frac{1}{\pi^2} \int d^4 a \delta (a^2-1)=\frac{1}{2\pi^2} \int d\psi d\theta d\phi \sin^2 2\psi \sin \theta
\end{equation}
where $a$ is defined through $U_0=a_0 +i \vec{a} \cdot \vec{\sigma}$. 

\subsection{Left-current correlator}
If all light quarks are in the $\epsilon$-regime and for $N_l=2$ we obtain
\begin{equation}
{\cal C}^{(\epsilon)l}_{J}(t)|_{N_l=2} =\theta(t) \frac{a^2}{8}  \exp\left(-\Delta M^{(\epsilon)} t\right)
\Bigg[1+\frac{3}{4} \frac{1}{(F L)^2 } \left(H(t, L, T) +g_\pi^2 H'(t,L,T) \right) \Bigg].
\label{eq:hmchpt}
\end{equation}
where 
\ba
H(t, L, T) &\equiv & L^2 \left(\frac{T}{L^3}h_1\left(\frac{t}{T}  \right)+\frac{1}{L^3}\sum_{\vec{p}\neq 0}P(t,|\vec{p}|)-\overline{G}(0,0)\right),\nonumber\\
H'(t, L, T)&\equiv & {1 \over L}  \sum_{\vec{p}\neq 0}
\left(P(t,|\vec{p}|)- P(0, |\vec{p}|)\right).
\label{eq:H}
\ea
and 
\ba
\label{dmasse}
\Delta M^{(\epsilon)} \equiv    \frac{3g_\pi^2}{8 F^2 L^3}.
\ea
The functions $h_1$ and $P$ are defined in Eq. (\ref{h1}), (\ref{def_P}), while the propagator $\overline{G}(0,0)$ is given in Eq. (\ref{beta1}).
\\
This expression contains no divergences in dimensional regularisation.
It is interesting to stress the fact that the zero-mode integrals that contribute to various diagrams, nicely cancel in the sum of all contributions. In particular this means that the current correlator loses any dependence on quark masses close enough to the chiral limit, which also means no dependence on the topological sector. \\
This result may be used to predict the behaviour of a correlator of left currents with the quantum numbers of the $B$ meson, in a finite volume such that the $u$ and $d$ quarks are in the $\epsilon$-regime.

\begin{figure}[htbp]
\begin{center}
\includegraphics[width=9cm]{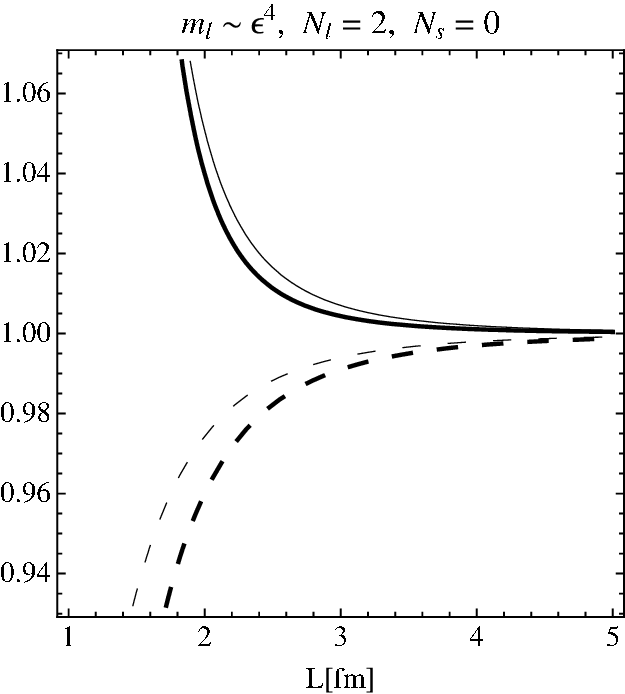} 
\caption{ 
Ratio of ${\cal C}_J^{(\epsilon)l}(t=1 fm)$ at fixed volume normalised to the $\infty$ volume result as a function of $L$ for two boxes with $T=L$ (solid) and $T=2 L$ (dashed), and for $g_\pi=0.44$ (thick lines) \protect\cite{Becirevic:2009yb} and $g_\pi=0$ (thin lines). We have fixed $\Sigma=(250\;{\rm MeV})^3, F=90$ MeV.  }
\label{fig:epsilon}
\end{center}
\end{figure}

In Fig.~\ref{fig:epsilon}, we show the ratio of the finite-volume to infinite volume correlator at $t=1$ fm as a function of the volume for two boxes and two values of $g_\pi$ ($g_\pi=0$ and  $g_\pi=0.44$, as recently computed on the lattice by \cite{Becirevic:2009yb}).  Corrections are ${\cal O}(3-4\%)$ at 2 fm, and the dependence on $g_\pi$ is mild. 

\begin{figure}[htbp]
\begin{center}
\includegraphics[width=9cm]{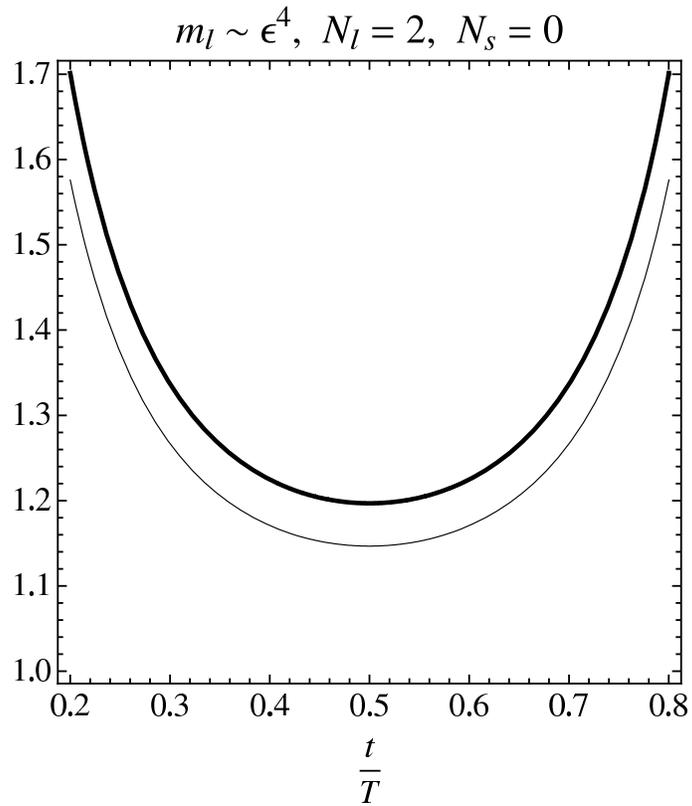} 
\caption{ 
$4 {\cal C}_J^{(\epsilon)l}(t)/(a^2 \exp(-\Delta M^{(\epsilon)} t))$ as a function of $t/T$  for $T=L=2$ fm, and for $g_\pi=0.44$ (thick line) \protect\cite{Becirevic:2009yb} and $g_\pi=0$ (thin line).}
\label{fig:epsilonx0}
\end{center}
\end{figure}

In Fig.~\ref{fig:epsilonx0}, we show the time dependence of the correlator after factoring out the $\exp(- \Delta M^{(\epsilon)} t)$ (we will see later in Sec. \ref{sec:lat} that in any real fit to lattice data, $\Delta M^{(\epsilon)}$ would renormalise the static energy $E_{stat}$). 

\subsection{Pseudoscalar density correlator}

For the pseudoscalar density, the result at the LO for arbitrary $N_l$ is found to be:
\begin{equation}
{\cal C}^{l}_{P}(t)=\frac{a^2}{8} \theta(t)\left[\langle(U_0+U_0^\dagger)_{ll}\rangle +2\right].  
\label{pseudocorHM}
\end{equation}

\section{HMChPT in mixed-regime}

\subsection{Setup}

In order to keep into account the effects due to the strange quark in heavy-light systems it is convenient to apply the power counting introduced in \cite{mixed} and reviewed in section \ref{sec:mixed}. To implement it in HQET, at least in the $N_l=2$, $N_s=1$ specific case does not require more technology than the one introduced in the previous section. \\
In practice all the steps described in the previous section must be applied again to the parametrisation given in Eq. (\ref{parU}).

\subsection{Left-current correlator}

In this case the $SU(3)$ vectorial symmetry is explicitly broken by the fact that 2 light quarks have mass $m_l$ lying in the $\epsilon$-regime while the one playing the role of the strange has a mass $m_s$ in the $p$-regime. This explains why the result is different depending on which light quark appears in the external line.
As before we first consider the case in which $l=1,2$. This result  represents the correlator of a left current with the quantum numbers of a $B$ or a $B^*$ in the context of 2+1 light flavours. We obtained:
\ba
\label{mixres12}
{\cal C}_{J}^{(m)1,2}(t)&=&\theta(t)\frac{a^2}{8}  \exp\left(-\Delta M^{(m_1)} t\right) \Bigg\{ 1+ 2 m_s\eta_3         + \nonumber \\
&&+\frac{1}{2F^2 L^2}\left[\frac{3}{2} \left( H(t,L,T) +g_\pi^2 H'(t,L,T) \right) \right.+\nonumber\\
&&+\frac{1}{L}\sum_{\vec{p}} \left(\left(P(t,M_{s\vec{p}})-P(0,M_{s\vec{p}})\right)   \left(1+{g_\pi^2 \vec{p}^2 \over M_{s\vec{p}}^2}\right)  \right) \nonumber\\
&& \left. +\frac{1}{6}\frac{1}{L}\sum_{\vec{p}} \left(\left(P(t,M_{\eta_s \vec{p}})-P(0,M_{\eta_{s}\vec{p}})\right) \left(1+{g_\pi^2 \vec{p}^2 \over M_{\eta_s\vec{p}}^2}\right)  \right) \right] \Bigg\}\,\,,
\ea
where $M_{\eta_s}$ has been defined in Eq. (\ref{etas}) and
\ba
\label{dmassm1}
\Delta M^{(m_1)} \equiv 2 m_s \sigma'_1 +  {g_\pi^2\over 4 F^2 L^3}  \left( \frac{3}{2} + \sum_{\vec{p}} \frac{M_s^2}{M_{s\vec{p}}^2} + {1\over 6}  \sum_{\vec{p}} \frac{M_{\eta_s}^2}{M_{\eta_s\vec{p}}^2}\right).
\ea
 This correlator will be matched with the predictions from the mixed ChPT, case B.

\begin{figure}[htbp]
\begin{center}
\includegraphics[width=9cm]{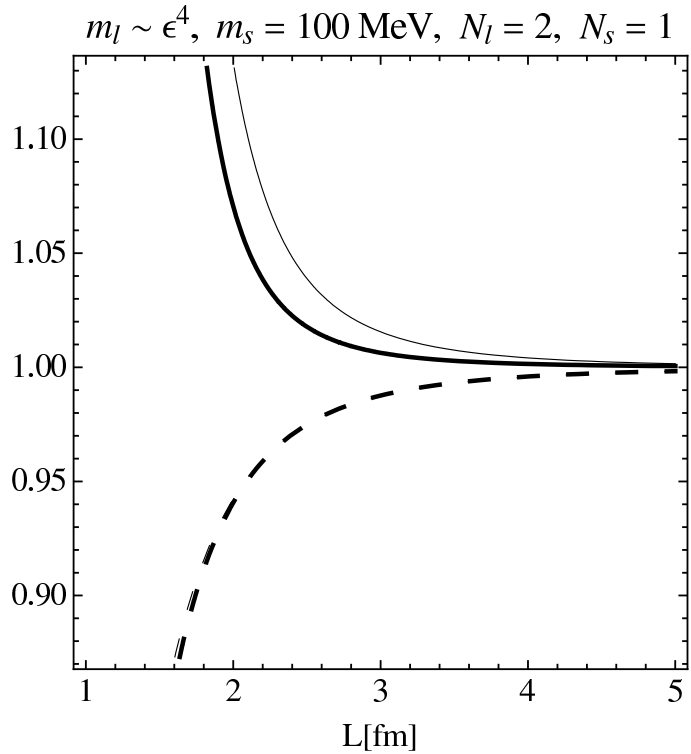} 
\caption{ 
Ratio of ${\cal C}_J^{(m)1}(t=1 fm)$ at fixed volume normalised to the $\infty$ volume result as a function of $L$ for two boxes with $T=L$ (solid) and $T=2 L$ (dashed), and for $g_\pi=0.44$ (thick lines) \protect\cite{Becirevic:2009yb} and $g_\pi=0$ (thin lines). The fact that the curves for $T=2L$ for $g_\pi=0$ or 0.44 nearly coincide is accidental.}
\label{fig:mixed}
\end{center}
\end{figure}

In Fig.~\ref{fig:mixed}, we show the ratio of the finite-volume to infinite volume correlator as a function of the volume for two boxes and two values of $g_\pi$ in the mixed regime. We have set $N_l=2$ and $N_s=1$. The corrections are qualitatively 
similar to those in the $\epsilon$-regime and quantitatively a bit larger. 

\begin{figure}[htbp]
\begin{center}
\includegraphics[width=9cm]{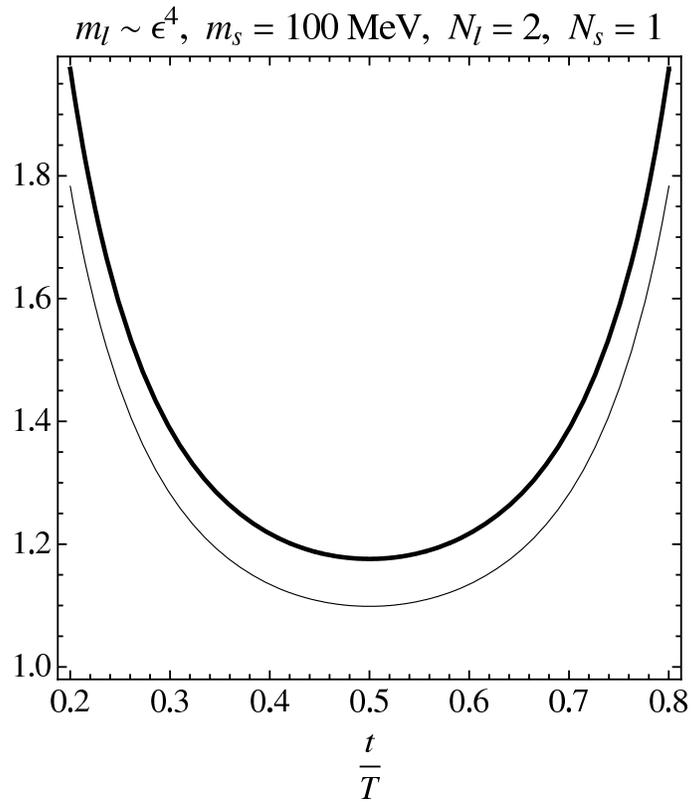} 
\caption{ 
$4 {\cal C}_J^{(m)1}(t)/(a^2 \exp(-\Delta M^{(m_1)} t))$ as a function of $t/T$  for $T=L=2$ fm, and for $g_\pi=0.44$ (thick line) \protect\cite{Becirevic:2009yb} and $g_\pi=0$ (thin line).}
\label{fig:mixedx0}
\end{center}
\end{figure}

In Fig.~\ref{fig:mixedx0}, we show the time dependence of the correlator after factoring out the $\exp(- \Delta M^{(m_1)} t)$. 

Another reason why HMChPT is useful is to predict the relation between observables related to the $B$ ($B^*$) and the $B_s$ ($B_s^*$). So we add for completeness also the results representing the correlator ${\cal C}^{(m)3}_{J}(t)$ of two left currents with the quantum numbers of a $B_s$ (or a $B_s^*$). In this case we obtain
\ba
\label{mixres3}
{\cal C}^{(m)3}_{J}(t)&=&\theta(t) \frac{a^2}{8} \exp\left(-\Delta M^{(m_3)} t\right)\Bigg\{ 1+2m_s(\eta_0+\eta_3)        \nonumber \\
&&+\frac{1}{F^2 L^2}\left[\frac{1}{L}\sum_{\vec{p}} \left(\left(P(t,M_{s\vec{p}})-P(0,M_{s\vec{p}})\right)   \left(1+{g_\pi^2 \vec{p}^2 \over M_{s\vec{p}}^2}\right)  \right) \right. \nonumber\\
&& \left.+\frac{1}{3}\frac{1}{L}\sum_{\vec{p}} \left(\left(P(t,M_{\eta_s \vec{p}})-P(0,M_{\eta_{s}\vec{p}})\right) \left(1+{g_\pi^2 \vec{p}^2 \over M_{\eta_s\vec{p}}^2}\right)   \right) \right] \Bigg\}\,\,,
\ea
where
\ba
\label{dmassm3}
\Delta M^{(m_3)} \equiv 2 m_s (\sigma_1+ \sigma'_1) +  {g_\pi^2\over 2 F^2 L^3}  \left( \sum_{\vec{p}} \frac{M_s^2}{M_{s\vec{p}}^2} + {1\over 3}  \sum_{\vec{p}} \frac{M_{\eta_s}^2}{M_{\eta_s\vec{p}}^2}\right).
\ea
Note that also in this case even though the various diagrams do depend on $m_l$, the final result does not.

%% file: matching.tex
\section{Matching of HMChPT and ChPT}
\label{sec:comp}

Dominant finite-size effects in QCD are due to pion dynamics, since these are the lightest degrees of freedom. It is therefore expected that the finite-size scaling of heavy-light systems does not 
depend on the large energy scales related to the heavy quarks, ie. $M_{hh}$ or $M_{hl}$. This must be the case as long as those scales are significantly larger than $L^{-1}$. Whether these scales are much larger also than the QCD scale so that the static limit (HQET) is a good approximation, or not, 
should not matter a priori for the finite-size scaling properties, because the volume dependence arises from the propagation of  the light degrees of freedom. \\
The leading finite volume effects are therefore expected to come from the fact that the heavy meson can emit and absorb a pion. The probability for this to happen can however depend on the heavy mass scale. 
Close enough to the chiral limit, the masses of pseudoscalar mesons are suppressed by the spontaneous breaking of chiral symmetry, so, for example, we do not need to include the vector mesons in the effective theory, because they are much heavier and decouple. On the other hand, in the limit $m_h \to \infty$ pseudoscalar and vector mesons are degenerate, because the interaction between quark and antiquark inside the meson becomes spin independent, so they both need to be considered in HMChPT. The presence of heavy-light vector resonances can modify the finite volume effects indirectly by inducing unsuppressed  contributions to pion/heavy-light meson scattering. We will see that indeed the finite-size corrections in HMChPT and mixed ChPT match up to corrections proportional to $g^2_\pi$. 

Real $c$ and $b$ quarks are somewhere in between these limits, where no effective description is very accurate. We may ask whether it is possible to give a description of finite size effects in this intermediate regime. In particular, there might be other resonances to consider\cite{Becirevic:2007dg}. 
 Using general arguments it was shown in  \cite{Nussinov:1999sx} that the pseudoscalar meson remains the lightest state for every value of the quark masses. Moreover if the heavy quark is in the non relativistic regime, we can say that the axial and scalar mesons (made of the same quark-antiquark couple) are heavier because they are in a higher angular momentum state. Experiments show that this peculiarity persists for heavy-light mesons whose heavy quark is a strange or a charm, the mass difference among the vector meson and the axial one being always of order of 400~MeV \cite{PdG2008}.\\
Finally the fact that exotic states may play a significant role is disfavoured by large $N_c$ arguments \cite{Witten:1979kh} saying that quark bilinears amplitude to produce them (like a $\overline{q}q\overline{q}q$) is suppressed.\\
To sum up it seems plausible to consider a scenario in which the current correlator has two channels, a pseudoscalar and a vectorial one. The vectorial one could be integrated out for quark masses that are small compared to $\Lambda_{QCD}$. Indeed it is known \cite{PdG2008} that while the $K$s weight approximately 500 MeV the $K^* $s weight approximately 900 MeV and while pions weight 140 MeV the $\rho$s weight 770 MeV.\\
However the vectorial channel becomes more and more relevant as the mass of the heavy quark grows, because the mass difference between pseudoscalar and vector mesons diminishes: if for $K$ mesons it is about 400 MeV, for $D$s it is about 150 MeV, while for $B$s it is only 50 MeV.

We consider now how the matching works in the two examples considered. Given any meson two-point function, the first point to realise is that a finite static limit is recovered after factorising out the leading $e^{- M |t|}$, where $M$ is the mass of the heavy meson and $t$ is the temporal separation between the two mesonic sources. 

\subsection{Pseudoscalar two-point function}

Let us start with the pseudoscalar correlator at LO, which is given in Eq. (\ref{pseudocorChPT}) for the mixed ChPT case and in Eq. 
(\ref{pseudocorHM}) for the HMChPT case. The first thing we observe is that the contribution of the zero modes, in particular the factor 
\begin{equation}
\left[\langle (U_0+U_0^\dagger)_{ll} \rangle +2\right]
\end{equation}
appear in both correlators. This shows that the zero modes contributions match in the two frameworks.

Moreover, if we use the expansion
\begin{eqnarray}
\lim_{M \rightarrow \infty} P(t, M^2) \rightarrow \theta(t) {e^{-M t}\over 2 M}  + {\mathcal O}\left( e^{- M T} \right), 
\end{eqnarray}
in Eq.~(\ref{pseudocorChPT}) we obtain
\begin{equation}
\lim_{M_h \rightarrow \infty} C_{P}(t)\rightarrow \frac{\Sigma^2}{4 F^2 M_h}  \theta(t) e^{-M_h t} \left[\langle  (U_0+U_0^\dagger)_{ll} \rangle +2\right] .
\end{equation}
After factorising out the exponential $e^{-M_h t}$ we find that also the time-dependence matches exactly the one predicted by the HMChPT in  
 Eq. (\ref{pseudocorHM}).
 The matching of the coefficient gives 
 
\begin{equation}
{a^2 \over 2} = {\Sigma^2 \over F^2 M_h} = F^2 M_h ({M_h \over m_h})^2 ,
\end{equation}
which  in the heavy quark mass limit $M_h \sim m_h$ is consistent with the definition in Eq.(\ref{def_a}), $a^2/2=F^2_P M_P$. 
Since in the static limit there is no time dependence at LO, we have that the ratio of correlation functions at different volumes $V_1$ and $V_2$ is given by
\begin{eqnarray}
{\mathcal{C}^l_P(t) |_{V_1} \over \mathcal{C}^l_P(t) |_{V_2}} =  { \langle (U_0 + U_0^\dagger )_{ll}\rangle  + 2|_{V_1} \over \langle (U_0 + U_0^\dagger )_{ll}\rangle  + 2 |_{V_2}}  \label{CPP_v1v2} 
\end{eqnarray}
in both regimes of the heavy quark mass.

\subsection{Left-current two-point function}

We consider the matching for the Dirac components $\mu=\nu=4$ for which we have the ChPT result. 

 What can be matched is the dependence of the correlators on the volume, that is $L$ and $T$ and the masses of the up, down and strange quarks, because these are explicit degrees of freedom in both effective theories. Moreover, since we only consider the static limit of HMChPT, we have to drop from the ChPT result those contributions that are suppressed by negative powers of $m_h$. We expect that the $\epsilon$ regime ($I=\epsilon$) HMChPT result should match to case $(A)$ in the mixed-regime computation, while the $I=m$ result in HMChPT should match case $(B)$. 
\begin{itemize}
\item
\emph{Case A}: \\
In order to match Eqs.~(\ref{eq:hmchpt}) and (\ref{eq:caj}),  the $L, T$ dependence must be the same in both cases. 
For the mixed ChPT framework, we split the contribution due to the heavy pions from the rest in Eqs. (\ref{FANLO}) and (\ref{MANLO}) and write:
\ba
F_{(A)}^2&=&  \overline{F}^2(m_h,\,N_l)- {1 \over 2} \left(N_l - {1\over N_l}\right)\overline{G}(0,0)+O(m_h^{-1}),\\
M_{(A)}^2&=& \overline{M}_h^2(m_h\,N_l)+O(m_h^{-1}).
\ea
$ \overline{F}$ and $\overline{M}_h$ have absorbed the dependence on the heavy quark mass. The static limit of the mixed ChPT case in Eq.~(\ref{eq:caj}) is, for $t>0$:
\begin{eqnarray}
{C^{(A)}_{J}(t)\over \exp(-\overline{M}_ht)} &=& {\overline{F}^2\overline{M}_h\over 4} \left[ 1   + {1 \over 2 F^2L^2} \left(N_l - {1\over N_l}\right) H(t,L,T) \right], 
\label{eq:deg}
\end{eqnarray}
where $H(t,L,T)$ is the function of Eq.~(\ref{eq:H}).

For $N_l=2$, the result is identical to the NLO prediction ${\cal C}^{(\epsilon)}_{J}(t)$ in HMChPT (Eq.~(\ref{eq:hmchpt})) with the following identifications:
\begin{eqnarray}
a=F_P\sqrt{2 M_P}=\overline{F}\sqrt{2 \overline{M}_h}, \qquad  g_\pi = 0 .
\label{identifications}
\end{eqnarray}
The fact that at NLO we have to put $g_{\pi}=0$ to match the two expressions reflects the fact that the vector meson is integrated out in the chiral theory. 
In HMChPT the vector and pseudoscalar are degenerate and therefore both are present. More generally we would expect that in the intermediate regime 
the finite size scaling of the current correlator behaves as
\begin{eqnarray}
{{\cal C}_J^{(\epsilon)}(t)|_{V_1}\over {\cal C}_J^{(\epsilon)}(t)|_{V_2}} &=& 1 + {1\over 2 F^2 L_1^2} \left( N_l - {1 \over N_l}\right) \left( H(t,L_1,T_1) + 
\alpha(t, L_1, T_1, m_h)\right)\\
& &- {1\over 2 F^2 L_2^2} \left( N_l - {1 \over N_l}\right) \left( H(t,L_2,T_2) + \alpha(t, L_2, T_2, m_h)\right)+...
\label{eq:all}
\end{eqnarray}
where $V_1= L_1^3 T_1$ and $V_2=L_2^3 T_2$ and
\begin{eqnarray}
&\lim_{m_h \rightarrow 0} \alpha(t,L,T,m_h) =0,\nonumber\\
&\lim_{m_h \rightarrow \infty} \alpha(t,L,T,m_h) = g_\pi^2 H'(t,L,T).   
\end{eqnarray}
In the intermediate region the function $\alpha$ is unknown. However it should be possible to compute it including the leading $1/m_h$ corrections in HMChPT or even in 
ChPT including the vector resonance, as a function of the vector meson mass and coupling. We will not consider these regimes in the present work. 
 Note however that for the value of $g_\pi$ obtained in a recent lattice computation in \cite{Becirevic:2009yb}, $g_\pi=0.44$, the contribution of the 
 term proportional to $g_\pi^2$ (i.e. the difference between the  thick and thin curves) in Figs.~\ref{fig:epsilon}-\ref{fig:mixedx0} is not too large, and should decrease with decreasing $m_h$.

\item \emph{Case B}: \\

We have to follow the same steps as above, but in addition to $L$ and $T$, we expect to reproduce also the dependence on $m_s$, up to $m_h^{-1}$ contributions. 
In the mixed ChPT framework, we rewrite Eqs. (\ref{FBNLO}) and (\ref{MBNLO}) as:
\begin{eqnarray}
F_{(B)}^2 &=&\overline{F}^2(m_h,\,N_l+N_s)-  {1 \over 2} \left(N_l - {1\over N_l}\right) \overline{G}(0,0)- {N_s \over 2 } G\left(0, M_{s}\right) +\\
& &-{N_s \over 2 N N_l} G\left(0, M_{\eta_s}\right) + 8 N_s M_{ss}^2 L_4,\nonumber\\
M_{(B)}^2&=&\overline{M}_{h}^2(m_h,\,N_l+N_s)-\frac{8M_{ss}^2M_{h}^2N_s(L_4-2L_6)}{F^2}\,\,.
\end{eqnarray}
The mixed correlator in the static limit can then be written as:
\begin{eqnarray}
{C^{(B)}_J(t) \over \exp(- \overline{M}_{h}t)} &=& {\overline{F}^2\overline{M_h}\over 4} \exp \left( \frac{4M_{ss}^2M_{h}N_s(L_4-2L_6)}{F^2}t\right) \nonumber \\
&&\hspace{-2cm} \left[ 
1  + \frac{4N_s M_{ss}^2}{F^2} (L_4+2L_6) +
{1 \over 2 L^2F^2} \left(N_l - {1 \over N_l}\right) H(t,L,T)  +
\right. \nonumber\\
&& \hspace{-2cm}
\left. +
 {1\over 2 L^3F^2} \left(  N_s\sum_{p} \left(P\left(t,M_{s\vec{p}}\right) -P\left(0,M_{s\vec{p}}\right) \right)+\frac{N_s}{N_lN}\sum_{p} \left(P\left(t,M_{\eta_s\vec{p}}\right) -P\left(0,M_{\eta_s\vec{p}}\right) \right) \right)\right] \nonumber.
\label{eq:nondeg}
\end{eqnarray}
One can check straightforwardly that, for $N_l=2$ and $N_s=1$, this coincides with the correlator ${\cal C}_J^{(m_1)1,2}(t)$ computed in HMChPT (Eq. (\ref{mixres12})) with the identifications:
\ba
a &=& F_P\sqrt{2 M_P}=\overline{F}\sqrt{2 \overline{M}_h}, \nonumber \\
g_\pi &=& 0, \nonumber \\
\eta_3^{(r)} & = & \frac{4\Sigma}{F^4}(L_4^{(r)}+2L_6^{(r)}), \nonumber \\
\sigma'_1 &=& - \frac{4\Sigma M_h}{F^4}(L_4^{(r)}-2L_6^{(r)}). 
\ea
Note that the above relations are among  renormalised quantities.  
Apart from some finite volume effects due to the sea $p$-regime quarks, which are exponentially suppressed, the volume dependence is identical to the one of Case A. So again we expect that for any value of $m_h$,  Eq.~(\ref{eq:all}) holds up to higher order chiral corrections and neglecting exponentially suppressed terms in $\exp(-M_{s} L)$.
\end{itemize}

\section{Finite-size scaling of heavy-light mesons in lattice QCD}
\label{sec:lat}

As we have seen above the matching of finite-size effects of heavy-light correlators in HMChPT and ChPT works as expected. We are  interested however in using these results to predict the finite-size scaling of these correlators  computed in lattice QCD. On the lattice, we can include a relativistic or static heavy quark.  In both cases we expect that for sufficiently large time separations:
\ba
C^{lat}_J(t) \equiv \sum_{\vec{x}} \langle J_\mu^a(x) J_\mu^a(0) \rangle_{lat} \simeq  {\cal C}^{ll}_J(t) \times {1\over 2M}\exp(- M t),
\ea
where $M$ is the lightest heavy-light meson mass $M_{hl}$ in the case of a relativistic heavy quark or the so-called static energy, $E_{stat} = M_{hl}-m_h$ in the lattice static limit.  \\
Note that the value of $E_{stat}$ is not predicted by HMChPT, however in general we can write:
\be
E_{stat} =E_{stat}^{(0)}+\Delta M
\ee
where $E_{stat}^{(0)}$ is the value the static energy would have in the chiral limit, while $\Delta M$ contains the chiral corrections that are predicted by HMChPT, that we have presented  for the various cases  considered, in Eqs.~ (\ref{dmassp}), (\ref{dmasse}), (\ref{dmassm1}) and (\ref{dmassm3}). \\
In practice this means that to fit a correlator evaluated with all the quarks in the $\epsilon$-regime using Eq.(\ref{eq:hmchpt}) one has to determine four parameters: $a$, $F$, $E_{stat}^{(0)}$ and $g_\pi$. It remains to be seen 
what the stability of such fits is in practice. The numerical challenge of extracting  signals over the noise when computing propagators of heavy mesons is well known. Recent proposals to improve the situation have been discussed in \cite{DellaMorte:2003mn, DellaMorte:2005yc}.

%% file: concl.tex
\section{Conclusions}\label{sec:conclu}

We have considered the finite-size scaling of heavy-light mesons, composed of a light quark in the $\epsilon$-regime. We have computed the left-current and pseudoscalar two-point functions in two limiting regimes of the heavy quark mass:
a  small heavy quark mass such that the heavy-light meson can be treated in the mixed-regime of ChPT, and the static limit where HMChPT can be applied.   We confirm  the naive expectation that the dominant finite volume effects 
are  induced by the emission/absorption of light pions,  and are to a large extent insensitive to the value of the heavy quark mass. These results can be useful for matching lattice QCD and ChPT or HMChPT in finite volumes  not sufficiently large compared with the Compton wavelength of the lighter pions. Our results can be used to consider also various partially-quenched situations.

%% file: appendices.tex
\appendix
\renewcommand{\theequation}{A-\arabic{equation}}
\setcounter{equation}{0}  
\section{Space time integrations in HMChPT}  

To obtain the charge correlators from the current ones, one has to integrate the current correlators over space. We report here the relevant results.

\subsection{The finite volume propagator in the rest frame}\label{appA1}

In this section we want to obtain the propagator of the heavy-light mesons in Euclidean space and at finite volume.
The propagator of HQET is obtained by writing the four-momentum of the heavy quark $p_{\mu} $ as: $p_{\mu}=m_h v_{\mu}+k_{\mu}$ and keeping only the leading term in the residual momentum $k_{\mu}$.
We consider here the rest frame, where $v=(0,0,0,i)$. 
For subtleties related to the Euclidean formulation for $\vec{v}\neq 0$ the reader can refer to \cite{Aglietti:1992in,Aglietti:1993hf}. 
In order to obtain the heavy quark propagator, one start from the Dirac quark propagator in coordinate space and we take the heavy quark limit, which is given by
\begin{equation}\label{propinfty}
S_\infty(x)=\frac{1}{(2\pi)^4}\int d^4p \frac{e^{ipx}(-ip_\mu\gamma_\mu +m_h) }{(p^2+m_h^2)}\rightarrow 
S_{\infty}^{hq}(t)=\frac{1+\gamma_4}{2} \int_{-\infty}^{+\infty}\frac{dp_4}{2\pi}\frac{e^{ip_4t}}{i(p_4-im_h)}=
\end{equation}
$$
=\left(\frac{1+\gamma_4}{2}\right)\theta(t)e^{-m_ht}.
$$
The projector $(1+\gamma_4)/2$ retains only the particle content of the heavy quark, and for this reason the propagation in (\ref{propinfty}) is forward in time.\\
In the effective theory the exponential is factorised, and the static propagator at infinite volume is \cite{Eichten:1989zv}
\begin{equation}
V_{\infty}(x)=\frac{1}{(2\pi)^4}\int d^4p\frac{e^{ipx}}{2i(p_4-i\epsilon)}=\frac{1}{2}\delta(\vec{x})\theta(t)\,\,.
\end{equation}
We now consider a finite box $V=L^3T$ with periodic boundary conditions.
Analogously to (\ref{propinfty}), the finite-volume Dirac propagator in the heavy quark limit is given by
\begin{equation}
S(x)\rightarrow    S^{hq}(t)=\frac{1+\gamma_4}{2}\frac{1}{T}\sum_{p_4}\frac{e^{ip_4t}}{i(p_4-im_h)}\,\,,
\end{equation}
that is, for $0\le t <T$,
\begin{equation}
\label{passage}
S^{hq}(t) =\frac{1+\gamma_4}{2} \left[ \theta(t)\frac{e^{-m_ht}}{1-e^{-m_hT}}\right].
\end{equation}
In the $m_h \to \infty$ limit, this reproduces the result of the infinite volume (\ref{propinfty}). 
Consequently, in the rest frame, the finite heavy volume propagator is
\begin{equation}\label{prop_static}
{V}(x)=\frac{1}{2}\delta(\vec{x})\theta(t),
\end{equation}
which is exactly the propagator we obtain from the kinetic term of the HMChPT Lagrangian, Eq. (\ref{lagrangian}).
The heavy propagator has the same form as in infinite volume: this is not surprising, since it describes a static particle, which is not sensitive to the presence of a finite box.

\subsection{Space integrals (p-regime)}\label{app_a2}

We present here the results for the integrals over space that are needed in finite-volume HMChPT  when the light quark is in the $p$-regime. $V(x)$ represents the static propagator, Eq. (\ref{prop_static}), while $G(x,M)$ is the pion propagator of Eq. (\ref{def_G}).
The function $P(t,M)$ is defined in Eq. (\ref{def_P}).
\begin{eqnarray}
A_1(t)&\equiv&\int d^3\vec{x}\,V(x)=\frac{1}{2}\theta(t) ;\\
A_2(t,M) &\equiv& \int d^3\vec{x}\,V(x)G(x,M)=\frac{\theta(t)}{2L^3}
\left[\sum_{\vec{p}}P(t,M_{\vec{p}})   \right];\\
A_{3;\alpha}(t,M) &\equiv&\int d^3\vec{x}\, d^4 z V(x-z)V(z)\partial_{x_\alpha}G(x-z,M) =\nonumber \\
&=& \delta_{\alpha 4}
\frac{\theta(t)}{4}\left[\frac{1}{L^3}\sum_{\vec{p}}P(t,M_{\vec{p}})-G(0,M)             \right];
\end{eqnarray}
\begin{eqnarray}
A_{4;\alpha\beta}(t,M)&\equiv &\int d^3\vec{x}\, d^4 z \, d^4 w 
V(x-z)V(z-w)V(w) \partial_{z_\alpha}\partial_{w_\beta} G(z-w,M) ;\\
A_{4;\alpha\beta}(t,M) &=&0  \quad \mbox{if }\alpha \ne \beta; \\
  A_{4;44}(t,M)&=& \frac{1}{8}\theta(t) \left[  G(0,M) -\frac{1}{L^3}\sum_{\vec{p}}P(t,M_{\vec{p}})  \right];\\
\sum_\alpha{A}_{4;\alpha \alpha}(t,M)&=&-\frac{M^2}{8L^3} \theta(t) \sum_{\vec{p}}\frac{1}{M_{\vec{p}}^2}\left[ \frac{t}{2}+P(t,M_{\vec{p}})- P(0,M_{\vec{p}}) \right] ;
\end{eqnarray}
\begin{eqnarray}
A_5(t)&\equiv&\int d^3\vec{x}\, d^4 z V(x-z) V(z)=\frac{1}{4}t\theta(t);\\
A_6(t)&\equiv& \int d^3\vec{x}\, d^4 z \,d^4 w V(x-z) V(z-w) V(w)=\frac{1}{16}t^2\theta(t).
\end{eqnarray}

\subsection{Space integrals ($\epsilon $-regime)}

In the $\epsilon $-regime the integrals to be computed are the same as above, with $G(x,M)$ substituted by $\overline{G}(x,0)$ defined in Eq. (\ref{def_Gbar}). We will denote the corresponding integrals by $\overline{A}_n(t)$ instead of $A_n(t,M)$. 
We have obtained:
\ba
\overline{A}_2(t)    &=&    \frac{\theta(t)}{2L^3}\left[Th_1\left(\frac{t}{T}\right)+\sum_{\vec{p}\neq 0}P(t,|\vec{p}|)   \right]; \\
\overline{A}_{3\alpha}(t)   &=&   \delta_{\alpha 4}
\frac{\theta(t)}{4}\left[\frac{T}{L^3}h_1\left(\frac{t}{T}  \right)+\frac{1}{L^3}\sum_{\vec{p}\neq 0}P(t,|\vec{p}|)-     \overline{G}(0,0)    \right] ;  \\
\overline{A}_{4;44}(t)   &=&   -\frac{1}{8}\theta(t)\left[\frac{T}{L^3}h_1\left(\frac{t}{T}\right) +\frac{1}{L^3}\sum_{\vec{p}\neq 0}P(t,|\vec{p}|)-\overline{G}(0,0)          \right] ; \\
\sum_\alpha\overline{A}_{4;\alpha\alpha}(t)&=&  -\frac{1}{16}\theta(t) \frac{t^2}{V};\\
\overline{A}_{4;\alpha\beta}(t)   &=&   0 \quad \mbox{if }\alpha \ne \beta \,\,.
\ea
The function $h_1(t/T)$ is defined in Eq. (\ref{h1}).
Notice that no new integrals have to be considered in the mixed-regime case.

%% file: hqetF.bbl
\begin{thebibliography}{10}

\bibitem{Grinstein:1990mj}
B. Grinstein,
\newblock Nucl. Phys. B339 (1990) 253.

\bibitem{Eichten:1989zv}
E. Eichten and B.R. Hill,
\newblock Phys. Lett. B234 (1990) 511.

\bibitem{Georgi:1990um}
H. Georgi,
\newblock Phys. Lett. B240 (1990) 447.

\bibitem{Gamiz:2008iv}
E. Gamiz,
\newblock (2008), 0811.4146.

\bibitem{Gasser:1986vb}
J. Gasser and H. Leutwyler,
\newblock Phys. Lett. B184 (1987) 83.

\bibitem{Gasser:1987ah}
J. Gasser and H. Leutwyler,
\newblock Phys. Lett. B188 (1987) 477.

\bibitem{Gasser:1987zq}
J. Gasser and H. Leutwyler,
\newblock Nucl. Phys. B307 (1988) 763.

\bibitem{mixed}
F. Bernardoni and P. Hern\'andez,
\newblock JHEP 10 (2007) 033, 0707.3887.

\bibitem{GL2}
J. Gasser and H. Leutwyler,
\newblock Nucl. Phys. B250 (1985) 465.

\bibitem{mixed2}
F. Bernardoni et~al.,
\newblock JHEP 10 (2008) 008, 0808.1986.

\bibitem{Damgaard:2001js}
P.H. Damgaard et~al.,
\newblock Nucl. Phys. B629 (2002) 445, hep-lat/0112016.

\bibitem{Damgaard:2007ep}
P.H. Damgaard and H. Fukaya,
\newblock Nucl. Phys. B793 (2008) 160, 0707.3740.

\bibitem{Kanzieper:2002ix}
E. Kanzieper,
\newblock Phys. Rev. Lett. 89 (2002) 250201, cond-mat/0207745.

\bibitem{Splittorff:2002eb}
K. Splittorff and J.J.M. Verbaarschot,
\newblock Phys. Rev. Lett. 90 (2003) 041601, cond-mat/0209594.

\bibitem{Hasenfratz:1989pk}
P. Hasenfratz and H. Leutwyler,
\newblock Nucl. Phys. B343 (1990) 241.

\bibitem{Burdman:1992gh}
G. Burdman and J.F. Donoghue,
\newblock Phys. Lett. B280 (1992) 287.

\bibitem{Wise:1992hn}
M.B. Wise,
\newblock Phys. Rev. D45 (1992) 2188.

\bibitem{Yan:1992gz}
T.M. Yan et~al.,
\newblock Phys. Rev. D46 (1992) 1148.

\bibitem{Arndt:2004bg}
D. Arndt and C.J.D. Lin,
\newblock Phys. Rev. D70 (2004) 014503, hep-lat/0403012.

\bibitem{Smigielski:2007pe}
B. Smigielski and J. Wasem,
\newblock Phys. Rev. D76 (2007) 074503, 0706.3731.

\bibitem{Manohar:2000dt}
A.V. Manohar and M.B. Wise,
\newblock Camb. Monogr. Part. Phys. Nucl. Phys. Cosmol. 10 (2000) 1.

\bibitem{Casalbuoni:1996pg}
R. Casalbuoni et~al.,
\newblock Phys. Rept. 281 (1997) 145, hep-ph/9605342.

\bibitem{Aglietti:1992in}
U. Aglietti, M. Crisafulli and M. Masetti,
\newblock Phys. Lett. B294 (1992) 281.

\bibitem{Aglietti:1993hf}
U. Aglietti,
\newblock Nucl. Phys. B421 (1994) 191, hep-ph/9304274.

\bibitem{deDivitiis:1998kj}
UKQCD, G.M. de~Divitiis et~al.,
\newblock JHEP 10 (1998) 010, hep-lat/9807032.

\bibitem{Abada:2002xe}
A. Abada et~al.,
\newblock Phys. Rev. D66 (2002) 074504, hep-ph/0206237.

\bibitem{Abada:2003un}
A. Abada et~al.,
\newblock JHEP 02 (2004) 016, hep-lat/0310050.

\bibitem{Negishi:2006sc}
S. Negishi, H. Matsufuru and T. Onogi,
\newblock Prog. Theor. Phys. 117 (2007) 275, hep-lat/0612029.

\bibitem{Ohki:2008py}
H. Ohki, H. Matsufuru and T. Onogi,
\newblock Phys. Rev. D77 (2008) 094509, 0802.1563.

\bibitem{Becirevic:2009yb}
D. Becirevic et~al.,
\newblock (2009), 0905.3355.

\bibitem{Boyd:1994pa}
C.G. Boyd and B. Grinstein,
\newblock Nucl. Phys. B442 (1995) 205, hep-ph/9402340.

\bibitem{Goity:1992tp}
J.L. Goity,
\newblock Phys. Rev. D46 (1992) 3929, hep-ph/9206230.

\bibitem{Hansen:1990un}
F.C. Hansen,
\newblock Nucl. Phys. B345 (1990) 685.

\bibitem{Becirevic:2007dg}
D. Becirevic, S. Fajfer and J.F. Kamenik,
\newblock PoS LAT2007 (2007) 063, 0710.3496.

\bibitem{Nussinov:1999sx}
S. Nussinov and M.A. Lampert,
\newblock Phys. Rept. 362 (2002) 193, hep-ph/9911532.

\bibitem{PdG2008}
Particle Data Group, C. Amsler et~al.,
\newblock Phys. Lett. B667 (2008) 1.

\bibitem{Witten:1979kh}
E. Witten,
\newblock Nucl. Phys. B160 (1979) 57.

\bibitem{DellaMorte:2003mn}
ALPHA, M. Della~Morte et~al.,
\newblock Phys. Lett. B581 (2004) 93, hep-lat/0307021.

\bibitem{DellaMorte:2005yc}
M. Della~Morte, A. Shindler and R. Sommer,
\newblock JHEP 08 (2005) 051, hep-lat/0506008.

\end{thebibliography}
